\documentclass[twocolumn,showpacs,preprintnumbers,amsmath,amssymb,floatfix]{revtex4}
\input{epsf}

\usepackage{graphicx}% Include figure files
\usepackage{dcolumn}% Align table columns on decimal point
\usepackage{bm}% bold math

\begin{document}

%\preprint{APS/123-QED}

\title{Quantum communication between trapped ions through a dissipative environment}
\author{H. T. Ng and S. Bose}
\affiliation{{Department of Physics and Astronomy, University
College London, Gower Street, London WC1E 6BT, United Kingdom}}
\date{\today}

\begin{abstract}
We study two trapped ions coupled to the axial phonon modes of a
one-dimensional Coulomb crystal. This system is formally equivalent
to the ``two spin-boson'' model. We propose a scheme to dynamically
generate a maximally entangled state of two ions within a
decoherence-free subspace. Here the phononic environment of the
trapped ions, whatever its temperature and number of modes, serves
as the entangling bus. The efficient production of the pure singlet
state can be exploited to perform short-ranged quantum communication
which is essential in building up a large-scale quantum computer.
\end{abstract}


\maketitle

Spin-boson model plays an important role in studying open quantum
systems in physics and chemistry \cite{Leggett,Breuer}.  Recently,
Porras {\it et al.} \cite{Porras} have proposed to use a trapped ion
coupled to a set of collective modes of a one-dimensional (1D)
Coulomb crystal such that the spin-boson model can be realized and
studied precisely. This model can be physically realized by shining
a laser on an ion in which the two internal states are coupled to a
traveling wave \cite{Porras}.  The axial motional modes of 1D
Coulomb crystal act as a phonon bath and provide the Ohmic spectral
density. This paves the way for the experimental studies of the
Ohmic spin-boson model in the low as well as high temperature
regimes.

In fact, sophisticated techniques have been demonstrated in the
manipulation of trapped ions such as cooling the ions to the
motional ground state and detecting the state of the ions
\cite{Leibfried}. Besides, Coulomb crystals of ion gases have been
observed in Paul \cite{Diedrich} and Penning traps \cite{Itano} and
storage rings \cite{Schatz}. A few dozens of ions separated by
several micrometers in a crystal form have been observed. This
provides a promising ground to investigate the spin-boson model with
trapped ions.

\begin{figure}[ht]
\includegraphics[height=2cm]{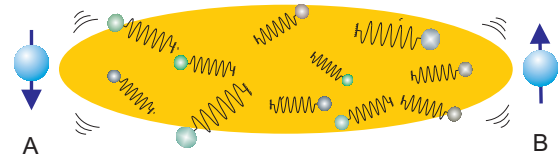}
\caption{ \label{fig1} (Color online) Two ions are coupled to a set
of axial phonon modes of the Coulomb chian which is acted as a
harmonic bath.}
\end{figure}

The work of this paper will be based on the ``two spin-boson'' model
which can be implemented in a 1D Coulomb crystal. It is indeed a
generalization of the scheme by Porras {\it et al.} \cite{Porras} to
study the spin-boson model. The two ions are now considered to
couple to a set of axial phonon modes of the chain. This phonon bath
plays the role of a common dissipative environment for the two ions
(see Fig. \ref{fig1}). Nevertheless, the decoherence can be greatly
canceled provided that the two ions are prepared in the opposite
spin polarizations and their separation is sufficiently short
compared to the correlation length of the Coulomb chain
\cite{Massimo}. It is the so-called subdecoherent space
\cite{Massimo,Duan} or decoherence-free subspace (DFS) \cite{Lidar}
in which the decoherence can be completely quenched. This is very
useful in protecting qubits if they are encoded as logical qubits in
two physical qubits \cite{Kielpinski}.

Some studies of the generation of mixed-state entanglement between
the two spins in a common harmonic bath have been performed recently
\cite{Braun}. In this paper, we show that the phonon bath can
mediate maximal entanglement between the two ions within the DFS.
This means that a pure maximally entangled state, with long
coherence times, can be generated even in a noisy environment. This
robust entangled state could be very useful in the ion-trap quantum
computing, particularly on the issue of its scalability. Indeed, a
considerable amount of effort have been devoted to physical
realization of a scalable ion-trap quantum computer \cite{Leibfried}
since the first proposal of the quantum computer with trapped ions
\cite{Cirac}.

We propose to use the ion chain to perform short-ranged quantum
communication which is an important element of building up a
large-scale quantum computer. According to the blueprint of building
a large-scale quantum computer, a quantum computer composed of a
number of quantum registers is envisaged \cite{Kielpinski}. Each
quantum register is connected through a common quantum data bus. The
quantum gates can be performed in the individual quantum registers
and the different quantum registers can communicate with each other
through some quantum channels.  Typically, this quantum channel will
be optical, but alternatives, such as physically transporting the
otherwise stationary qubits \cite{Rowe} or using information
propagation in a chain of qubits \cite{Bose} are worth studying as
they enable one to avoid the issue of interfacing different types of
physical systems.

\begin{figure}[ht]
\includegraphics[height=6cm]{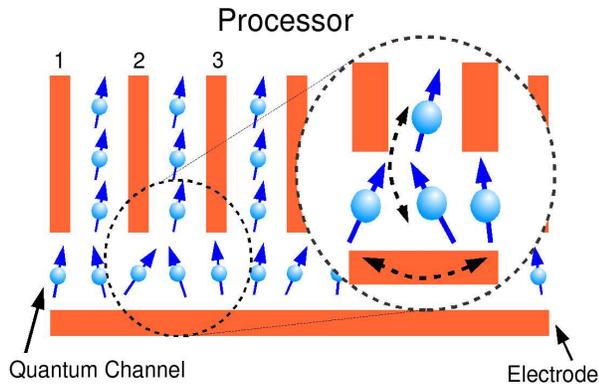}
\caption{ \label{fig2} (Color online) The quantum registers are
placed in different zones. The different quantum registers are
connected by a chain of ions which is served as a quantum channel.
An enlarged diagram shows that the quantum state of ions in the
quantum register can be transferred to an ion in the chain (dashed
arrow indication). The quantum information is then transmitted
through the channel.}
\end{figure}

We consider a situation in which ions in distinct quantum registers
are arranged in different zones and they are interconnected via a
chain of ions as shown in Fig. \ref{fig2}. The Coulomb ion chain can
thus be used as a quantum data-bus between quantum registers. For
example, the quantum state of any ion in the quantum register can be
transmitted to an ion in the chain with transverse phonon modes
\cite{Zhu}. The quantum information can then be teleported
\cite{Bennett2} through the ion chain using the entanglement between
two ions of the chain developed in accordance to the scheme of this
paper. In this way, the quantum communication between the different
quantum registers can be accomplished. Bearing in mind that the
quantum information has been encoded within the DFS, there is no
further requirement of entanglement distillation \cite{Bennett} and
a high fidelity of the teleportation can be achieved.

We consider $N$ ions arranged in a 1D Coulumb chain. The $N$ ions
are confined in a linear trap and interact with each other via the
Coulumb repulsion.  The trapping and the Coulumb potentials are of
the form \cite{Porras0,Morigi,Porras}
\begin{eqnarray}
V_{\rm
trap}&=&\frac{1}{2}{m}\sum^N_{i=1}(\omega^2_xx^2_i+\omega^2_yy^2_i+\omega^2_zz^2_i),\nonumber\\
V_{\rm
Coul}&=&\sum^{N}_{i>j}\frac{e^2}{\sqrt{(x_i-x_j)^2+(y_i-y_j)^2+(z_i-z_j)^2}},\nonumber
\end{eqnarray}
where $\omega_\alpha$ are the trapping frequencies in the direction
$\alpha=x,y,z$; $m$ and $e$ are the mass and the charge of each ion
respectively.  The ions form a linear chain along the $z-$direction
for $\omega_{x,y}\gg\omega_z$.  The motion of $N$ ions in 1D chain
is described by collective modes and the dispersion relation is
subject to the trapping condition.  The approximate eigenvalues of
the axial modes can be found as $\omega_n=\omega_z\sqrt{n(n+1)/2}$
for $N\gg{1}$ \cite{Morigi}. This approximation is indeed very good
for the low axial modes even in a chain of ten ions \cite{Morigi}.

The two ions are illuminated by two laser fields individually, the
ions interact with the traveling wave \cite{Porras,Leibfried}. The
internal states of two ions in the chain are then coupled to the
axial vibrations of the entire chain.   The Hamiltonian of the ions
and the phonon bath read as $(\hbar=1)$,
\begin{eqnarray}
H_{\rm
ion}&=&\frac{\omega}{2}\sum^2_{j=1}\sigma^j_z,\\
H_{B}&=&\sum_n\omega_nb^\dag_n{b}_n,
\end{eqnarray}
where $\omega$ and $\sigma^j_z$ are the energy splitting and Pauli
operator; $\omega_n$ and $b_n$ are the frequency and the
annihilation operator of the $n-$th phonon mode.  The Hamiltonian,
in a rotating frame with the laser frequency $\omega_L$, is written
as \cite{Leibfried}
\begin{equation}
H_{L}=\sum^2_{j=1}\frac{\Omega}{2}(\sigma^j_+e^{i\tilde{k}Z_j}+\sigma^j_-e^{-i\tilde{k}Z_j}),
\end{equation}
where $\Omega$ is the laser Rabi frequency, $\tilde{k}$ is the wave
number of the laser,
$Z_j=\sum_n\tilde{z}_n(b^\dag_ne^{ikz_j'}+b_ne^{-ikz_j'})$ is the
position operator of the chain.  The parameters $k$ and
$\tilde{z}_n$ are $2\pi{l}/L$ and $1/\sqrt{2m\omega_n}$
respectively, for $l=0,1,2,\ldots,N-1$, $L=Na$ is the length of
chain and $a$ is the separation between two neighboring ions
\cite{Morigi}.

By applying a unitary transformation
$U=e^{-i\tilde{k}(Z_1\sigma^1_{z}+Z_2\sigma^2_z)/2}$ \cite{Porras},
the total Hamiltonian is rewritten as
\begin{eqnarray}
\label{sbHam}
H&=&\sum^2_{j=1}\Big(\frac{\Delta}{2}\sigma^j_x+\frac{\omega_0}{2}{\sigma^j_z}\Big)+\sum_n{\omega_n}b^\dag_n{b}_n\nonumber\\
&&+\sum^2_{j=1}\sum_n\frac{\sigma^j_z}{2}(g^j_nb^\dag_n+g^{j*}_nb_n),
\end{eqnarray}
where $\Delta=\Omega$, $\omega_0=\omega-\omega_L$ and
$g^j_n=i\tilde{k}\tilde{z}_n\omega_ne^{ikz_j'}$.

The two spin-boson model can be solved exactly if the parameter
$\Delta$ is zero \cite{Massimo,Breuer}.  In order to get more
insight, we first study the Hamiltonian $H_0$ in which we set
$\Delta=0$ in Eq. (\ref{sbHam}). We apply a canonical transformation
$e^S=\exp[{\sum^2_{j=1}\sum_n\sigma^j_z({g^j_n}b^\dag_n-{g^{j*}_n}b_n)}/2{\omega_n}]$
to the Hamiltonian $H_0$ \cite{Leggett}. In the rotating frame, the
Hamiltonian is written as
\begin{equation}
\tilde{H}_0=\frac{\omega_0}{2}\sum^2_{j=1}\sigma^j_z+\sum_n\Big(\omega_nb^\dag_nb_n-\lambda\sigma^1_z\sigma^2_z
-\frac{|g_n|^2}{\omega_n}\Big),
\end{equation}
where $\lambda=\sum_n|g_n|^2\cos(kr)/\omega_n$ and $r=z_1'-z_2'$.
Here we consider the separation between the two qubits is
sufficiently short compared to the correlation length of the bath so
that the two qubits are effectively coupled to the common bath
\cite{Massimo}. Roughly speaking, the correlation length of the
Coulomb chain is about the chain length $L$ if only the low-lying
excited modes are involved.  Hence, we can impose a condition that
the separation $r$ must be much less than the length of the chain
$L$. If this condition is satisfied, the two ions will ``feel'' to
interact with the same harmonic bath.

The bath can be characterized by the spectral density function
$J(\omega)=\sum_k|g_k|^2\delta(\omega-\omega_j)$.  For a linear
Coulomb chain, in the low-lying excitation regime, the spectral
density $J(\omega)$ has been shown to be Ohmic in the case that the
ion interact with the traveling wave. If the ions are equally space
with a distance $a$, then the spectral density has the form
\cite{Porras}:
\begin{eqnarray}
\label{SD1} J(\omega)&=&\eta\omega{e^{-\omega/\omega_c}},
\end{eqnarray}
where $\eta=\tilde{k}^2/2m\omega_z{\nu}$ and
$\nu=(3e^2/m\omega^2_za^3)^{1/2}$.

We have introduced the cut-off frequency $\omega_c$ to the spectral
density in Eq. (\ref{SD1}) since it is legitimate to perform the
adiabatically elimination of the high frequency modes compared to
the tunneling strength $\Delta$ \cite{Leggett}. Nevertheless, it is
required that the cut-off frequency $\omega_c$ is much greater than
the laser strength $\Delta$ and the thermal energy $k_BT/\hbar$
\cite{Leggett}.  We also presume that this cut-off frequency
$\omega_c$ is low enough so that only the low-lying excitations are
involved in the bath dynamics.

Now we study the dynamics of trapped ions with a thermal phonon
bath. We consider that the trapped ions and the bath are separable
initially, i.e., $\rho_{\rm T}(0)=\rho(0){\otimes}\rho_B$, $\rho(0)$
and $\rho_B$ are the density matrix of the qubits and the thermal
bath respectively. The thermal bath is in equilibrium and its
density matrix is given by $\rho_B=e^{-\beta{H_B}}/Z_B$, where
$\beta=1/k_BT$, $k_B$ and $T$ are the Boltzmann constant and the
temperature respectively, and $Z_B={\rm Tr}e^{-\beta{H_B}}$ is the
partition function.

The reduced density matrix of the two qubits can be obtained by
tracing out the system of bath, i.e., $\rho(t)={\rm Tr}_B[{\rho_{\rm
T}}(t)]$. The reduced matrix of the two qubits is spanned by the
basis:
$\{|1\rangle=|11\rangle,|2\rangle=|10\rangle,|3\rangle=|01\rangle,|4\rangle=|00\rangle\}$.
Since the population number is conserved for $\Delta=0$,  the
evolution of the diagonal elements of the density matrix is
constant: $\rho_{ii}(t)=\rho_{ii}(0)$.  The matrix elements satisfy
$\rho_{ij}=\rho^*_{ji}$.  The non-diagonal matrix element decays as
\cite{Massimo}
\begin{eqnarray}
\rho_{12}(t)=\rho_{12}(0)e^{-i\phi_--\Gamma},~~~~\rho_{13}(t)=\rho_{13}(0)e^{-i\phi_--\Gamma},\nonumber\\
\rho_{24}(t)=\rho_{24}(0)e^{-i\phi_+-\Gamma},~~~~\rho_{34}(t)=\rho_{34}(0)e^{-i\phi_+-\Gamma},\nonumber\\
\end{eqnarray}
where
\begin{eqnarray}
\phi_{\pm}&=&\omega_0t{\pm}2\lambda{t}\nonumber\\
&&{\pm}\int^{\infty}_0\frac{J(\omega)}{\omega^2}\coth\Big(\frac{\omega}{2k_BT}\Big)\sin\omega{t}d\omega,\\
\Gamma&=&\int^{\infty}_0\frac{J(\omega)}{\omega^2}\coth\Big(\frac{\omega}{2k_BT}\Big)(1-\cos\omega{t})d\omega.
\end{eqnarray}
The states under the effect of the collective decoherence read as
\cite{Massimo}
\begin{eqnarray}
\rho_{23}(t)&=&\rho_{23}(0)e^{-2\Gamma_-},\\
\rho_{14}(t)&=&\rho_{14}(0)e^{-2i\omega_0{t}-2\Gamma_+},
\end{eqnarray}
where
\begin{equation}
\Gamma_{\pm}=\int^{\infty}_0\frac{J(\omega)}{\omega^2}\coth\Big(\frac{\omega}{2k_BT}\Big)(1-\cos\omega{t})[1\pm\cos(kr)]d\omega.
\end{equation}

The decay rates of the qubits are dramatically changed due to the
collective decoherence.  For the states $\rho_{14}$ and $\rho_{41}$,
the decay process is greatly enhanced with the decay rate
$\Gamma_{+}$.  In contrast, the decay rate of the states $\rho_{23}$
and $\rho_{32}$ is largely reduced.  In the limit of zero separation
$r$, the states $|01\rangle$ and $|10\rangle$ are protected within
the DFS \cite{Duan}. Physically speaking, the notion of DFS holds if
the decay rate of this subspace is comparable or even longer than
the natural coherence times of the ion.  In fact, we can estimate
explicitly the coherence lifetime of this subspace
$\sim\Gamma^{-1}_-$ which is roughly equal to
$(\eta{k_BTt})^{-1}(\omega_cr/\omega_zL)^{-2}$ \cite{remark}.
Therefore, the long coherence lifetime can be achieved for the low
temperature $T$ and the small separation $r{\ll}L$ .

Now we investigate the effect of the two local laser fields on the
two qubits for a nonzero tunneling parameter $\Delta$.  The
transformed Hamiltonian $\tilde{V}=e^{S}Ve^{-S}$ can be obtained
\begin{eqnarray}
\tilde{V}&=&\sum^2_{j=1}\frac{\Delta}{2}(\sigma^j_+B^\dag_j+B_j\sigma^j_-),
\end{eqnarray}
where $B_j=\exp[\sum_n(g^j_nb^\dag_n-g^{j*}_nb_n)/2\omega_n]$ and
$B^\dag_1B_2\approx{1}$ for a small separation $r$.

We consider the laser field strength $\Delta$ is much weaker than
the spin-bath coupling $\lambda$.  This enables us to derive an
effective Hamiltonian within the DFS based on the second-order
perturbation theory.  The matrix elements of the effective
Hamiltonian can be represented as
\begin{eqnarray}
(\tilde{H}_{\rm
eff})_{mn}&=&-\sum_l\frac{\tilde{V}_{ml}\tilde{V}_{ln}}{E^{(0)}_l-(E^{(0)}_m+E^{(0)}_n)/2},
\end{eqnarray}
where $m$ and $n$ represent the basis of unperturbed states:
$\{|01\rangle,|10\rangle\}$ and $l$ denotes the intermediate states
$\{|11\rangle,|00\rangle\}$; $E^{(0)}_{l}$ are the eigenenergies of
$\tilde{H}_0$.

The effective Hamiltonian can be obtained as \cite{Ng}
\begin{eqnarray}
\tilde{H}_{\rm eff}&=&-\kappa(J_+J_-+J_-J_+),
\end{eqnarray}
where $\kappa=\lambda\Delta^2/2(\omega^2_0-4\lambda^2)$ and
$J_{\pm}=\sum^2_{i=1}\sigma^i_{\pm}$. The dynamics within the DFS is
governed by this effective Hamiltonian.   The state with the initial
state $|10\rangle$ evolves
\begin{equation}
|\Psi(t)\rangle=e^{2i\kappa{t}}[\cos2\kappa{t}|10\rangle+i\sin2\kappa{t}|01\rangle].
\end{equation}
The two qubits become entangled when the two local laser fields are
turned on.  At the time $t^*=\pi/8\kappa$, up to a global phase
factor $e^{i\pi/4}$, the quantum state is given by,
\begin{eqnarray}
|\Psi(t^*)\rangle&=&\frac{1}{\sqrt{2}}(|10\rangle+i|01\rangle).
\end{eqnarray}
An ideal entangled state is then generated dynamically. Here the
spin-coupling strength is about $10^7$ to $10^8$ Hz
($\lambda\varpropto\sum_n\omega_n$) if the trapping frequency
$\omega_z$ can attain up to several MHz. Therefore, the speed of
entanglement formation can be estimated around $10^4$ to $10^5$ Hz
where the laser Rabi frequency $\Omega$ have to be much smaller than
$\lambda$.

We notice that the entanglement generation  shares the same spirit
of the S{\o}rensen-M{\o}lmer entangling gate in which the
entanglement gate is implemented by coupling to the virtual motional
states of a single centre-of-mass (CM) mode \cite{Sorensen}.  It is
indeed inevitable to couple a number of phonon modes in performing
the gate operation as considered by Jonathan and Plenio
\cite{Jonathan}. However, we have studied a more general scenario by
consideration of position-dependent couplings to the phonon bath in
which the collective decoherence effect sets in.  The entanglement
can be efficiently generated within the DFS whereas the
superposition of the states $|11\rangle$ and $|00\rangle$ is very
fragile.

Having discussed the generation of the entangled state, we proceed
to study the quantum communication between two trapped ions. This
can be used to transfer quantum information between different
quantum registers by implementation of the quantum teleportation
protocol \cite{Bennett}. For instance, we consider to transmit the
quantum information from the ion $i$ to the ion $j$. We first
generate the entanglement between the ions $j$ and $k$.  The ion $k$
is next to the ion $i$. Then, we perform the Bell-state measurement
between the ions $i$ and $k$. The quantum teleportation can thus be
accomplished by sending the measurement result to the ion $j$
through the classical communication. It is noteworthy that the range
of quantum state transfer should be short in our scheme. However,
this can be resolved by repeating the quantum teleportation between
two nearby ions several times until the quantum information being
transferred to that distant ion.

Additionally, the ion chain can form a cluster state
\cite{Raussendorf} by sequential generation of entangled states
between different pairs of ions, with the entangling of each pair
taking place according to the scheme of this paper, which will be
useful for measurement-based quantum computing.

In conclusion, we have investigated the two ions coupled to axial
vibration modes of the 1D Coulumb chain. This can be shown to be
equivalent to the ``two spin-boson'' model. We show that the
decoherence-free entanglement of two nearby ions can be dynamically
generated. It can be applied to perform short-ranged quantum
communication. The present result could also benefit quantum
information processing in solid-state based systems which can be
described by the spin-boson model such as the Josephson charge
qubits of a Cooper-pair box \cite{Nakamura} and semiconductor double
quantum dots \cite{Hayashi}.

We thank Andrew Fisher, Anna Sanpera and Dimitris Angelakis for
useful discussions. H. T. Ng is supported by the Quantum Information
Processing IRC (QIPIRC) (GR/S82176/01). S. Bose also thanks the
Engineering and Physical Sciences Research Council (EPSRC) UK for an
Advanced Research Fellowship and the Royal Society and the Wolfson
Foundation.


\begin{thebibliography}{99}
\bibitem{Leggett}
A. J. Leggett {\it et al.}, Rev. Mod. Phys. {\bf 59}, 1 (1987).

\bibitem{Breuer}
H.-P. Breuer and F. Petruccione, {\it The Theory of Open Quantum
Systems} (Oxford University Press, Oxford, 2002).


\bibitem{Porras}
D. Porras {\it et al.}, \textit{arXiv:0710.5145}.

\bibitem{Leibfried}
D. Leibfried {\it et al.}, Rev. Mod. Phys. {\bf 75}, 281 (2003).


\bibitem{Diedrich}
F. Diedrich {\it et al.}, Phys. Rev. Lett. {\bf 59}, 2931 (1987); R.
Bl\"{e}mel {\it et al.}, Nature (London) {\bf 334}, 309 (1998).

\bibitem{Itano}
W. M. Itano {\it et al.}, Science {\bf 279}, 686 (1998); T. B.
Mitchell {\it et al.}, Science {\bf 282}, 1290 (1998).


\bibitem{Schatz}
T. Sch\"{a}tz, U. Schramm and D. Habs, Nature (London) {\bf 412},
717 (2001).

\bibitem{Massimo}
G. Massimo {\it et al.}, Proc. R. Soc. Lond. A {\bf 452}, 567
(1996).

\bibitem{Duan}
L.-M. Duan and G.-C. Guo, Phys. Rev. Lett. {\bf 79}, 1953 (1997)

\bibitem{Lidar}
D. A. Lidar {\it et al.}, Phys. Rev. Lett. {\bf 81}, 2594 (1998).


\bibitem{Kielpinski}
D. Kielpinski, C. Monroe and D. J. Wineland, Nature {\bf 417}, 709
(2002).


\bibitem{Braun}
D. Braun, Phys. Rev. Lett., {\bf 89}, 277901 (2002).


\bibitem{Cirac}
J. I. Cirac and P. Zoller, Phys. Rev. Lett. {\bf 74}, 4091 (1995).




\bibitem{Rowe}
M. A. Rowe {\it et al.}, Quantum Inf. Comput. {\bf 2} 257 (2002),
arXiv:quant-ph/0205094.

\bibitem{Bose}
S. Bose, Contempary Physics, {\bf 48}, 13 (2007).

\bibitem{Zhu}
S.-L. Zhu, C. Monroe and L.-M. Duan, Phys. Rev. Lett. {\bf 97},
050505 (2006).

\bibitem{Bennett2}
C. H. Bennett {\it et al.}, Phys. Rev. Lett. {\bf 70}, 1895 (1993).

\bibitem{Bennett}
C. H. Bennett {\it et al.}, Phys. Rev. Lett. {\bf 78}, 2031 (1997).



\bibitem{Porras0}
D. Porras and J. I. Cirac, Phys. Rev. Lett. {\bf 96}, 250501 (2006).

\bibitem{Morigi}
G. Morigi and S. Fishman, Phys. Rev. Lett. {\bf 93}, 170602 (2004);
G. Morigi and S. Fishman, Phys. Rev. E {\bf 70}, 066141 (2004).









\bibitem{Ng}
H. T. Ng, C. K. Law and P. T. Leung, Phys. Rev. A {\bf 68}, 013604
(2003).

\bibitem{remark}
The decay rate $\Gamma_-$ is bounded by $(k_cr)^2I$ for
$k_cr\ll{1}$,
$I=\int^{\infty}_0J(\omega)/\omega^2\coth(\omega/2k_BT)(1-\cos{\omega{t}})d\omega$
and the wave number $k_c\approx2\pi\omega_c/\omega_zL$ is chosen at
the cutoff frequency.  The integral $I$ can be found analytically
for the Ohmic spectral densdity
$J(\omega)=\eta{\omega}e^{-\omega/\omega_c}$. It is equal to
$I{\approx}(1/2)\ln(1+\omega^2_c{t}^2)+\ln[\sinh(\pi{k_B}Tt)/\pi{k_B}{T}t]$
\cite{Breuer,Massimo}. Here we have assumed that
$\omega_c{\gg}k_BT$.  For $t\gg{1/\pi{k_B}T}$,
$I\approx{\pi{k_BT}t}$ is linearly proportional to time $t$.








\bibitem{Sorensen}
A.  S{\o}rensen and K. M{\o}lmer, Phys. Rev. Lett. {\bf 82}, 1971
(1999).

\bibitem{Jonathan}
D. Jonathan and M. B. Plenio, Phys. Rev. Lett. {\bf 87}, 127901
(2001).



\bibitem{Raussendorf}
R. Raussendorf and H. J. Briegel, Phys. Rev. Lett. {\bf 86}, 5188
(2001).




\bibitem{Nakamura}
Y. Nakamura, Yu. A. Pashkin and J. S. Tsai, Nature {\bf 398}, 786
(1999); Y. Makhlin, G. Sch\"{o}n and A. Shnirman, Rev. Mod. Phys.
{\bf 73}, 357 (2001).

\bibitem{Hayashi}
T. Hayashi {\it et al.}, Phys. Rev. Lett. {\bf 91}, 226804 (2003).

\end{thebibliography}
\end{document}